\documentclass[preprint,showpacs,preprintnumbers,amsmath,amssymb]{revtex4}

\usepackage{graphicx}
\usepackage{dcolumn}
\usepackage{bm}

\begin{document}


\title{Emergence of Kondo-assisted Néel order in a Kondo necklace model}


\author{Hironori Yamaguchi$^{1,2}$, Shunsuke C. Furuya$^{3,4}$, Yu Tominaga$^{1}$, Takanori Kida$^{5}$, Koji Araki$^{6}$, and Masayuki Hagiwara$^{5}$}
\affiliation{
$^1$Department of Physics, Osaka Metropolitan University, Osaka 558-8531, Japan\\
$^2$Innovative Quantum Material Center (IQMC), Osaka Metropolitan University, Osaka 558-8531, Japan\\
$^3$Department of Liberal Arts, Saitama Medical University, Moroyama, Saitama 350-0495, Japan\\
$^4$Institute for Solid State Physics, the University of Tokyo, Chiba 277-8581, Japan\\
$^5$Center for Advanced High Magnetic Field Science (AHMF), Graduate School of Science, Osaka University, Osaka 560-0043, Japan\\
$^6$Department of Applied Physics, National Defense Academy, Kanagawa 239-8686, Japan
}

\date{\today}

\begin{abstract}
The interplay between Kondo screening and magnetic order has long been a central issue in the physics of strongly correlated systems. While the Kondo effect has traditionally been understood to suppress magnetism through the formation of local singlets, recent studies suggest that Kondo interactions may enhance magnetic order under certain conditions. 
However, these scenarios often rely on complex electronic structures, including orbital and charge degrees of freedom, making the essential mechanisms difficult to isolate.
Here we report the realization of a spin-(1/2,1) Kondo necklace model in a Ni-based complex--a minimal spin-only analogue of the Kondo lattice that isolates quantum spin correlations by eliminating charge degrees of freedom. 
Thermodynamic measurements identify a magnetic phase transition and a field-induced quantum phase transition. 
Perturbative analysis reveals that the Kondo coupling mediates effective antiferromagnetic interactions between the spin-1 sites, stabilizing the Néel order across the entire chain. 
Our results establish a universal boundary in Kondo physics, where coupling to spin-1/2 moments yields singlets, but to spin-1 and higher stabilizes magnetic order.
\end{abstract}

\maketitle 
The Kondo effect, a hallmark of strongly correlated electron systems, results from antiferromagnetic (AF) coupling between localized moments and conduction electrons~\cite{Kondo}. 
This interaction forms spin singlets at low temperatures, leading to emergent behaviors such as heavy fermion states, non-Fermi liquid properties, and unconventional superconductivity~\cite{spcond1,spcond2,spcond3,spcond4,nonFermi}. 
The Kondo lattice model and its associated Doniach phase diagram have been central in describing the balance between Kondo screening and magnetic order via RKKY interactions~\cite{Doniach,Doni1, Doni2, Doni3}.
Conventionally, increasing Kondo coupling is thought to suppress magnetic ordering, an idea foundational to understanding quantum criticality in $f$-electron systems. 
However, recent studies on heavy fermion and valence-fluctuating compounds suggest that Kondo interactions may, under certain conditions, coexist with or even enhance magnetic order~\cite{Kondo1,Kondo2,Kondo3}. 
These interpretations often rely on complex mechanisms involving orbital degrees of freedom, valence fluctuations, or multipolar interactions, making the fundamental physics difficult to isolate.

To clarify these roles, spin-only models such as the Kondo necklace offer a valuable simplification by removing charge degrees of freedom~\cite{Doniach}. 
In its basic form, the model couples a spin-1/2 chain to local spins via Kondo-like exchange, capturing the interplay between singlet formation and magnetic correlations. 
Most studies have focused on the spin-1/2 model, where strong Kondo coupling typically leads to a nonmagnetic singlet ground state with a finite excitation gap~\cite{1DKN1,1DKN2,1DKN3,1DKN4,1DKN5,1DKN6,myKondo}.
The behavior becomes less clear when the local spins have higher spin values. 
The Kondo necklace model with spin-1 introduces richer spin dynamics and anisotropic effects, raising the question of whether Kondo coupling continues to suppress magnetism or might instead mediate effective interactions that stabilize magnetic order. 
Despite its conceptual importance, this regime remains unexplored both theoretically and experimentally.
Meanwhile, recent theoretical discussions in heavy-fermion systems have proposed Kondo-enhanced ordering, but these often rest on complex and indirect evidence.
A clear demonstration of Kondo-assisted magnetic order in a minimal, spin-only framework is still lacking. 
Realizing such behavior in the spin-(1/2,1) Kondo necklace model would provide compelling evidence for an alternative route to magnetism grounded purely in quantum spin correlations—a direction yet to be pursued in detail.
Moreover, this model permits tuning toward exotic gapless symmetry-protected topological phases~\cite{SPT}, providing a versatile framework for exploring entangled quantum states.

In this work, we realize a spin-(1/2,1) Kondo necklace model in a $[$Ni($p$-Py-V-$p$-F)(H$_2$O)$_5$$]$$\cdot$2NO$_3$ ($p$-Py-V-$p$-F = 3-(4-pyridinyl)-1-(4-fluorophenyl)-5-phenylverdazyl,) using a radical-based compound. 
The magnetic properties are well explained by the Kondo necklace model.
Comprehensive thermodynamic and ESR measurements reveals field-induced suppression of magnetic order and a critical field at which the spin-1 moments decouple from the spin-1/2 chain.
Perturbative analysis reveals that Kondo coupling induces effective AF interactions between the spin-1 sites, stabilizing the Néel order that propagates through the system.
The analysis also predicts a field-induced quantum phase transition, supporting the experimentally observed behavior.

\begin{figure*}[t]
\begin{center}
\includegraphics[width=30pc]{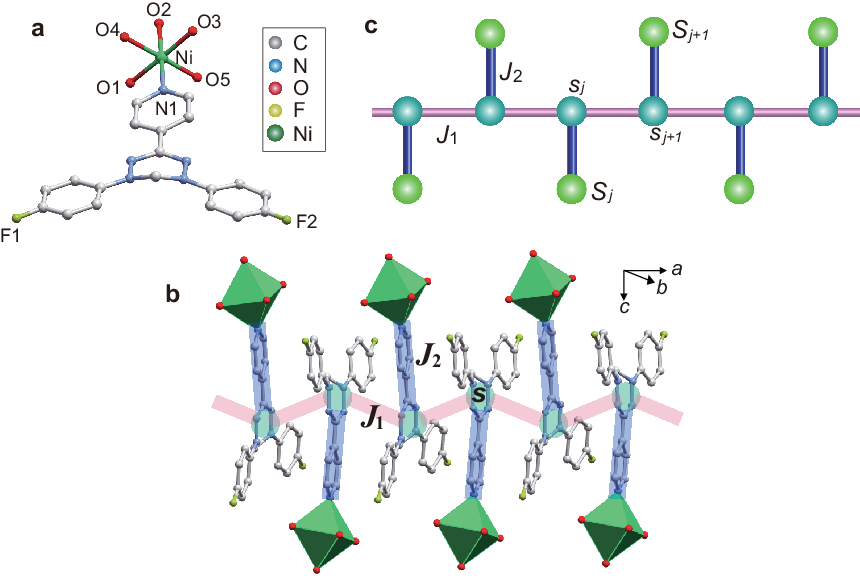}
\caption{\textbf{Crystal structure and Kondo necklace model of $[$Ni($p$-Py-V-$p$-F)(H$_2$O)$_5$$]$$\cdot$2NO$_3$.}
\textbf{a}, Molecular structure of Ni($p$-Py-V-$p$-F)(H$_2$O)$_5$. 
\textbf{b}, Crystal structure forming the Kondo necklace model along the $a$ axis. 
Hydrogen atoms are excluded to enhance clarity. 
The green nodes represent the spin-1/2 of the radicals.
The thick lines represent exchange interactions.
\textbf{c},Spin-(1/2,1) Kondo necklace model comprising intermolecular $J_{\rm{1}}$ and intramolecular $J_{\rm{2}}$. 
$\boldsymbol{s}$ and $\boldsymbol{S}$ denote the spins on the radical and Ni$^{2+}$, respectively. 
}
\end{center}
\end{figure*}

\begin{figure*}
\begin{center}
\includegraphics[width=35pc]{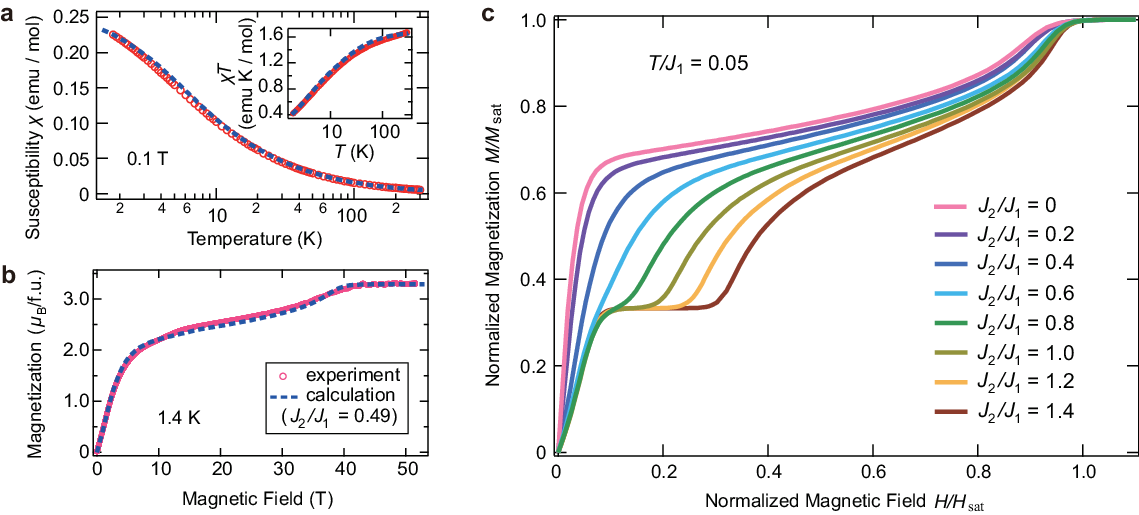}
\caption{\textbf{Magnetic behavior of $[$Ni($p$-Py-V-$p$-F)(H$_2$O)$_5$$]$$\cdot$2NO$_3$.}
\textbf{a},Temperature dependence of magnetic susceptibility ($\chi$ = $M/H$) of $[$Ni($p$-Py-V-$p$-F)(H$_2$O)$_5$$]$$\cdot$2NO$_3$ at 0.1 T. 
The inset shows corrresponding $\chi T$ values.
\textbf{b}, Magnetization curve of $[$Ni($p$-Py-V-$p$-F)(H$_2$O)$_5$$]$$\cdot$2NO$_3$ at 1.4 K.
The dashed lines represent the QMC results for the spin-(1/2,1) Kondo necklace model with $J_{1}/k_{\rm{B}}$ = 20.3 K and $J_{2}/k_{\rm{B}}$ = 9.9 K $J_{2}/J_{1}$ ($J_{2}/J_{1}$ = 0.49).
For the magnetization curve, a radical purity of 95 ${\%}$ is considered for the calculation.
\textbf{c}, Calculated magnetization curves at $T/J_{1}$ = 0.05 with the representative values of $J_{2}/J_{1}$. 
The magnetic moment and the magnetic field are normalized by the values at the saturation.
For $J_{2}/J_{1} \geq 1.0$, a clear 1/3 plateau emerges, reflecting full polarization of the effective spin-1/2 within the $\boldsymbol{s}$-$\boldsymbol{S}$ dimer mediated by $J_{2}$.
Conversely, for $J_{2}/J_{1} \ll 1.0$, the magnetization exhibits a steep rise toward 2/3, consistent with full polarization of the Ni spins caused by the effective decoupling of $J_{2}$.
}
\end{center}
\end{figure*}

\begin{figure*}
\begin{center}
\includegraphics[width=40pc]{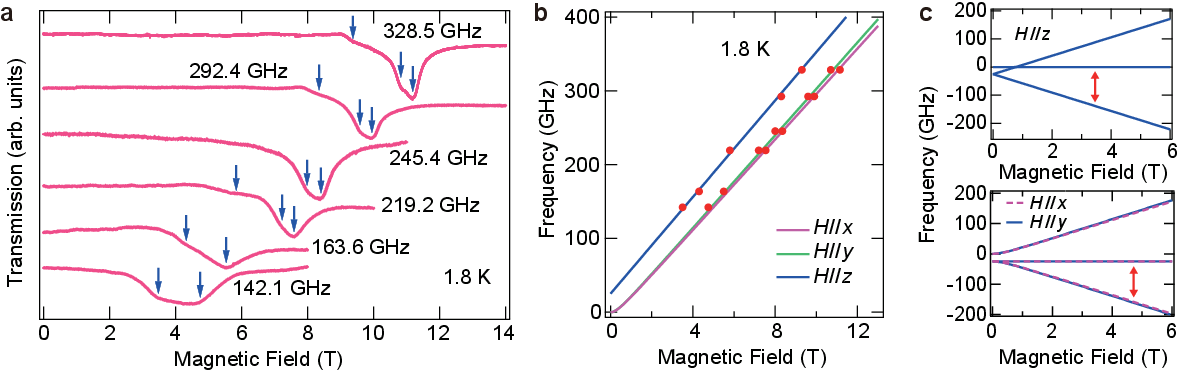}
\caption{\textbf{ESR of $[$Ni($p$-Py-V-$p$-F)(H$_2$O)$_5$$]$$\cdot$2NO$_3$.}
\textbf{a}, Frequency dependence of ESR absorption spectra of $[$Ni($p$-Py-V-$p$-F)(H$_2$O)$_5$$]$$\cdot$2NO$_3$ at 1.8 K. 
The arrows indicate the resonance fields.
\textbf{b}, Frequency-field plot of the resonance fields.
Solid lines indicate the calculated resonance modes of the spin-1 monomer along the principal axes, obtained using an on-site anisotropy of $D/k_{\rm{B}}$=$-$1.2 K and $g$-values of $g_x$=2.20, $g_y$=2.25, and $g_z$=2.34. 
\textbf{c}, Calculated energy branch of the the spin-1 monomer for $H$//$z$, $H$//$x$ and $H$//$y$.
Those for $H$//$x$ and $H$//$y$ are qualitatively equivalent.
Arrows indicate spin-allowed transitions from the ground state, which correspond to the resonance modes.
}
\end{center}
\end{figure*}

\begin{figure*}
\begin{center}
\includegraphics[width=35pc]{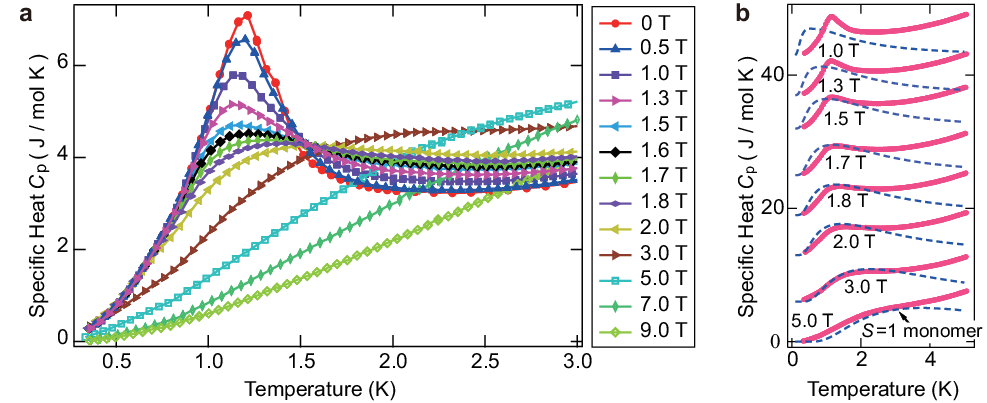}
\caption{\textbf{Specific heat of $[$Ni($p$-Py-V-$p$-F)(H$_2$O)$_5$$]$$\cdot$2NO$_3$.}
\textbf{a}, Temperature dependence of the specific heat $C_{\rm{p}}$ of $[$Ni($p$-Py-V-$p$-F)(H$_2$O)$_5$$]$$\cdot$2NO$_3$ at various magnetic fields for $H$//$b$, perpendicular to the chain direction.
The lines are guides for the eye. 
\textbf{b}, Low-temperature region of $C_{\rm{p}}$. 
For clarity, the values for 1.0, 1.3, 1.5, 1.7, 1.8, 2.0, and 3.0 T have been shifted up by 43, 37, 32, 25, 19, 13, and 6 J/ mol K,
respectively.
The broken lines represent the Schottky-type specific heat of the spin-1 monomer for $H$//$y$, obtained using an on-site anisotropy of $D/k_{\rm{B}}$=$-$1.2 K and $g_y$=2.25.
The experimental results approach the behavior of the spin-1 monomer with increasing fields, suggesting the critical field causing the decoupling of $\boldsymbol{S}$ is approximately 2 T.
The higher temperature deviations are considered to arise from the spin-1/2 chain and lattice contributions.
}
\end{center}
\end{figure*}

The molecular structure of Ni($p$-Py-V-$p$-F)(H$_2$O)$_5$ is shown in Fig. 1a, with the Ni atom surrounded by a pyridine in $p$-Py-V-$p$-F and five water (H$_2$O) ligand molecules, creating a six-coordinate environment.
In the $p$-Py-V-$p$-F molecule, the central ring comprises four nitrogen (N) atoms with a maximum spin density~\cite{TominagaCo}, leading to a localized spin system. 
Hence, the verdazyl radicals, $p$-Py-V-$p$-F, and Ni$^{2+}$ ions possess spin values of 1/2 and 1, respectively. 
The crystallographic parameters at 100 K are as follows: monoclinic, space group $P_{bca}$, $a$ =  8.37620(10)  $\rm{\AA}$, $b$ =  17.6025(3) $\rm{\AA}$, $c$ = 33.8343(6) $\rm{\AA}$, $V$ = 4988.60(14)  $\rm{\AA}^3$, $Z$ = 8 (Supplementary Table 1 and Table2).
The crystal structure is isomorphic to that of $[$Co($p$-Py-V)(H$_2$O)$_5$$]$$\cdot$2NO$_3$, forming a spin-1/2 Kondo necklace model with no chage degrees of freedom~\cite{myKondo}.
The MO calculations indicated predominant AF exchange interaction $J_{1}/k_{\rm{B}}$ = 14 K, forming a uniform spin-1/2 chain along the $a$ axis, as shown in Fig. 1b.
Each radical spin ($\boldsymbol{s}$) in the 1D chain is coupled with the Ni spin ($\boldsymbol{S}$) via intramolecular interactions, yielding a Kondo necklace model, as shown in Fig. 1b and 1c. 
The intramolecular interaction $J_{2}$ between the verdazyl radical and the Ni$^{2+}$ ion, corresponding to Kondo coupling, is expected to be of the order of 10 K, based on previous studies of the complex consisting of Ni and the verdazyl radical~\cite{TominagaNi}.
Moreover, no significant MO overlap was found between the 1D structures, enhancing the 1D nature of the Kondo necklace model (Supplementary Fig.1 and Note 1).

Figures 2a and 2b show the magnetic susceptibility ($\chi$ = $M/H$) at 0.1 T and magnetization curve at 1.4 K. 
We observe a paramagnetic-like increase in $\chi$ down to 1.8 K.
The magnetization curve also exhibits a paramagnetic-like behavior up to about 10 T, above which a nonlinear increase towards saturation at $\sim$40 T appears. 
Since the value of $\sim$2.2 $µ_{\rm{B}}$/f.u. at 10 T suggests field-polarization of the Ni spins, the higher-field nonlinear behavior is expected to originate from the radical spins forming the spin-1/2 AF chain.
Considering the results of MO calculations, the magnetic properties were investigated based on the spin-(1/2,1) Kondo necklace model composed of the AF interactions, $J_{1}$ and $J_{2}$.
We calculated $J_{2}/J_{1}$ dependence of the magnetization curve using the QMC method, as shown in Fig. 2c.
A clear 1/3-plateau appears for $J_{2}/J_{1}$ ${\geq}$ 1.0, indicating the full polarization of the effective spin-1/2 in the $\boldsymbol{s}$-$\boldsymbol{S}$ dimer via $J_{2}$.
As $J_{2}/J_{1}$ decreases, the plateau disappears, and a drastic increase towards 2/3 magnetization becomes apparent, suggesting the full polarization of the Ni spins due to the decoupling of $J_{2}$ as observed in the spin-1/2 anisotropic Kondo necklace model~\cite{myKondo}. 
Based on the parameter dependence, we obtained a good agreement between the experimental and calculated results, leading to the evaluations of $J_{1}/k_{\rm{B}}$ = 20.3 K and $J_{2}/k_{\rm{B}}$ = 9.9 K $J_{2}/J_{1}$ ($J_{2}/J_{1}$ = 0.49), as shown in Figs. 2a and 2b.
The quantitative agreement with the experimental data also confirms that possible F-site disorder has no appreciable effect on either $J_1$ or $J_2$.

We performed ESR measurements to examine the field-induced decoupling state of the Ni spin. 
Figures 3a shows the frequency dependences of the resonance signals at 1.8 K. 
Because the experiments were performed using the powder samples, the observed signals corresponded to the resonance fields for the external field parallel to the principal axes. 
We plotted the resonance fields in the frequency-field diagram, as shown in Fig. 3b.
Assuming the spin-1 monomer attributed to the decoupled $\boldsymbol{S}$, we consider the on-site anisotropy as $\mathcal {H} = D(S_{z})^2-\mu_{\rm{B}}\textbf{{\textit H}}\bf{\tilde  g}\textbf{{\textit S}}$, where $\mu_{B}$ is the Bohr magneton, and $\bf{\tilde  g}$ denotes the $g$-tensor.
The diagonal components for the principal axes of the $g$-tensor are $g_x$, $g_y$, and $g_z$ and the other components are zero.
Figure 3c show the energy levels calculated by using the evaluated parameters. 
The resonance modes at a sufficiently low temperature of 1.8 K correspond to the transitions indicated by the arrows in
the energy branches. 
As shown in Fig. 3b, we obtained a good agreement between the experimental and calculated results for the resonance modes using $D/k_{\rm{B}}$=$-$1.2 K, $g_x$=2.20, $g_y$=2.25, and $g_z$=2.34, demonstrating the field-induced decoupling of $\boldsymbol{S}$.


Figures 4a shows the temperature dependence of specific heat $C_{\rm{p}}$ for $H$//$b$, perpendicular to the chain direction. 
The lattice contributions are not subtracted from $C_{\rm{p}}$, but the magnetic contributions were expected to be dominant in the low-temperature regions considered here.
At zero-field, a sharp $\lambda$-type peak appears at $T_{\rm{N}}$ = 1.2 K, indicating a phase transition to an AF order. 
Note that the application of the magnetic field drastically reduces the peak value, resulting in a shift to a broad peak.
Considering the field-induced decoupling of $\boldsymbol{S}$, the broad peak corresponds to a Schottky behavior attributed to the energy gap of the spin-1 monomer.
The experimental results indeed approach the behavior of the spin-1 monomer with increasing fields, as shown in Fig. 4b.
Since the $C_{\rm{p}}$ of the spin-1 monomer for $H$//$y$ most closely reproduced the experimental results, the $b$ axis is assumed to be identical to the $y$ axis.
The higher temperature deviations are considered to arise from the spin-1/2 chain and lattice contributions.
The change in the peak shape with increasing fields suggests the critical field causing the decoupling of $\boldsymbol{S}$ is approximately 2 T.

We investigate the ground state of the spin-$(1,1/2)$ Kondo necklace model described by the spin Hamiltonian
\begin{equation}
\mathcal{H} = J_1 \sum_{j} \boldsymbol{s}_j \cdot \boldsymbol{s}_{j+1} + J_2 \sum_j \boldsymbol{s}_j \cdot \boldsymbol{S}_j + D \sum_j (S_j^z)^2.
\label{eq:Hamiltonian}
\end{equation}
In the limit $J_2 \ll J_1$, the dominant low-energy physics is governed by the spin-1/2 Heisenberg chain formed by the $\boldsymbol{s}$ spins. 
This system is known to exhibit a Tomonaga--Luttinger liquid (TLL) state, characterized by gapless excitations and algebraic spin correlations. 
To investigate the influence of the Kondo-type coupling $J_2$ on $\boldsymbol{S}$, we treat $J_2$ as a perturbative parameter (Supplementary Note 2). 
Then, the leading contribution arises from  excitations involving neighboring Kondo bonds, $(\boldsymbol{s}_j \cdot \boldsymbol{S}_j)(\boldsymbol{s}_{j+1} \cdot \boldsymbol{S}_{j+1})$, which generates an RKKY-type effective interaction between neighboring $\boldsymbol{S}$ sites, i.e., $J_{\text{eff}} \sum_j \boldsymbol{S}_j \cdot \boldsymbol{S}_{j+1}$.
The resulting effective coupling constant is given by
\begin{equation}
J_{\text{eff}} = \frac{2 J_2^2}{\pi^2 J_1} \sqrt{\frac{\pi}{2}} \approx J_1\left( \frac{J_2}{J_1} \right)^{2}{\times}0.254.
\label{eq:Jeff}
\end{equation}
This expression reflects the interplay between correlations in the $\boldsymbol{s}$ chain and the $\boldsymbol{s}$--$\boldsymbol{S}$ dimer. 
With the evaluated parameters, we obtain $J_{\text{eff}}/k_{\mathrm{B}} \approx 1.24~\mathrm{K}.$
This gives a ratio $D/J_{\text{eff}} \approx -0.97$, indicating that the single-ion anisotropy is strong enough to stabilize Néel order in the $\boldsymbol{S}$ sites~\cite{neel}. 
Once the $S_j$ spins develop Néel order, the Kondo coupling $J_2$ plays an important secondary role: it acts as a staggered longitudinal field on the $s_j$ spin chain. 
That is, the alternating $z$-component of the $S_j$ moments induces an effective staggered field on the $s_j$ sites through the Kondo interaction. 
As a result, the TLL state of the $\boldsymbol{s}$ chain becomes unstable and a finite staggered magnetization develops.
Therefore, the Néel order initially established on the $S_j$ sites propagates throughout the system, ultimately stabilizing a spontaneously symmetry-broken Néel order over the entire spin-(1,1/2) chain. 
This mechanism provides a consistent explanation for the magnetic order observed in the present study.
Importantly, this behavior is fundamentally distinct from the case where the Kondo coupling acts on $S=1/2$ local moments. 
In that situation, quantum correlations inevitably drive the system into a singlet ground state~\cite{1DKN1,1DKN2,1DKN3,1DKN4,1DKN5,1DKN6,myKondo}. 
Such singlet formation is unavoidable for $S=1/2$, and thus the emergence of Néel order through Kondo coupling is precluded. 
In sharp contrast, when the local moments are $S \geq 1$, the same Kondo mechanism can stabilize long-range AF order. 
Our present results therefore highlight a universal principle: the fate of Kondo-coupled systems is qualitatively determined by the size of the local spin, with $S=1/2$ favoring singlet liquid states and $S \geq 1$ supporting magnetic order.


We now analyze the excitation state associated with the Néel order. 
Assuming the Néel order of $\boldsymbol{S}$, we apply linear spin-wave theory to estimate the spin gap at zero magnetic field (Supplementary Note 3).
In this framework, the excitation gap $\Delta$ corresponds to the lowest spin-wave energy and is given by
\begin{equation}
\Delta = \sqrt{D^2 S^2 + 4 J_{\text{eff}} |D| S^2}.
\label{eq:gap}
\end{equation}
Substituting the parameters yields a zero-field spin gap of $\Delta / k_{\mathrm{B}} \approx 2.7~\mathrm{K}$.
We note that in strictly one-dimensional systems quantum fluctuations are strong, so the spin-wave approach is not quantitatively precise; here it is used only to provide a qualitative description and an appropriate energy scale.
For the present system, this temperature corresponds to the regime in which effective interchain couplings can be enhanced by thermally activated magnon excitations, leading to a phase transition to a long-range order~\cite{peak,morotaMn}. 
In this context, the energy scale of the observed phase transition temperature, $T_{\mathrm{N}} \sim 1.2~\mathrm{K}$, is consistent with the gap value derived from the effective spin-wave analysis.

Next we consider the effect of the applied magnetic field for $H//y$, which is the case for the specific heat measurements (Supplementary Note 4). 
In the linear spin-wave approximation, the excitation gap at $k=0$ increases with the magnetic field, while the gap at $k=\pi$ decreases. 
As in the zero-field Néel state, the canted Néel order of $\boldsymbol{S}$ is expected to induce a corresponding canted order in the $\boldsymbol{s}$ chain via the $J_2$ interaction. 
This canted Néel phase becomes unstable at the critical field $H_c$ where the magnon gap at $k = \pi$ vanishes, signaling a quantum phase transition.
The critical field is given by 
\begin{equation}
H_c = 4 J_{\text{eff}} S \sqrt{1 + \frac{|D|}{4 J_{\text{eff}}}},
\end{equation}
Using the evaluated parameters, we obtain $H_c \approx 3.7~\mathrm{T}$. 
Although this value somewhat exceeds the experimentally observed critical field of approximately $2~\mathrm{T}$, the discrepancy can be considered reasonably small, given that the linear spin-wave approximation does not incorporate intermagonon interactions and quantum fluctuations.
At $H_c$, the $\boldsymbol{S}$ spins are expected to decouple from the $\boldsymbol{s}$ chain, which is demonstrated in the experimental behaviors. 
The field-induced suppression of the Kondo coupling thus marks a phase transition from a magnetically ordered state to a quantum-disordered decoupled state. 
The consistency between the calculated spin gap, the critical field, and the thermodynamic measurements highlights the effectiveness of the perturbative approach in capturing the essential low-energy physics of the spin-(1,1/2) Kondo necklace model (Supplementary Note 5 and Table 3).
Such field-induced destabilization of magnetic order is further facilitated by the strong quantum fluctuations inherent in one dimensionality, which naturally render the Néel state fragile. 
Consequently, the disappearance of the order at $H_c$ and the emergence of TLL–like magnetization in the high-field regime are consistent with this one-dimensional character.
While our present data are well explained by a decoupling scenario, the precise nature of the high-field phase remains open. Inelastic neutron scattering could directly track the gap closing and subsequent excitations, while NMR could probe local fields and spin relaxation, providing further insight into the quantum correlations of the high-field phase.

\section*{Conclusion}
To summarize, we synthesized a Ni-based complex, [Ni($p$-Py-V-$p$-F)(H$_2$O)$_5$]$\cdot$2NO$_3$, that realizes a spin-$(1,1/2)$ Kondo necklace model.
Comprehensive thermodynamic and ESR measurements revealed field-induced suppression of magnetic order and a critical field at which the spin-1 moments decouple from the spin-1/2 chain.
Perturbative treatment of the Kondo coupling yields an effective AF interaction between spin-1 sites, which, together with single-ion anisotropy, stabilizes a Néel state.
This Néel order subsequently induces a symmetry breaking in the spin-1/2 chain via the Kondo coupling, resulting in a long-range magnetic order in the whole system.
The analysis also predicts a field-induced quantum phase transition, supporting the experimentally observed behavior.
Our results provide direct experimental evidence that Kondo coupling can enhance magnetic order even in a spin-only system—challenging the conventional Doniach scenario based on screening and magnetic suppression.
Kondo coupling inevitably drives $S=1/2$ moments into singlets, but for $S \geq 1$ it universally stabilizes long-range order—establishing a clear boundary in Kondo physics.
This finding not only broadens our fundamental understanding of Kondo lattice physics but also highlights a mechanism for quantum-state control via magnetic fields.

\section*{Methods}
The X-ray intensity data were collected using a Rigaku XtaLAB Synergy-S instrument.
Magnetization measurements were conducted using a commercial SQUID magnetometer (MPMS, Quantum Design).
High-field magnetization in pulsed magnetic fields was measured using a non-destructive pulse magnet.
Diamagnetic contribution calculated using Pascal's method was subtracted from the experimental data.
Specific heat measurements were performed using a commercial calorimeter (PPMS, Quantum Design). 
The ESR measurement was performed using a vector network analyzer (ABmm). 
Experiments were performed using powder samples.
For the specific heat measurements, we aligned plate-shaped microcrystals and applied the magnetic fields perpendicular to the crystal planes.
The molecular orbital (MO) calculations were performed using the UB3LYP method and a basis set of 6-31G~\cite{MOcal}. 
The quantum Monte Carlo (QMC) code is based on the directed loop algorithm in the stochastic series expansion representation~\cite{QMC2}. 
The calculations was performed for $N$ = 384 under the periodic boundary condition~\cite{ALPS,ALPS3}.
For the magnetic susceptibility, the calculated result was calibrated using the $g$ value of 2.22 as the average value.
For the magnetization curve, the calculated results were calibrated using the corresponding $g$ values, considering the predominant magnetic contributions~\cite{TominagaNi}.

We thank H. Shishido and Y. Shimura for valuable discussions.
A part of this work was performed under the interuniversity cooperative research program of the joint-research program of ISSP, the University of Tokyo.



\begin{thebibliography}{99}


\bibitem{Kondo}
Kondo, J. Resistance minimum in dilute magnetic alloys, $Progr.$ $Theoret.$ $Phys.$ \textbf{32}, 37 (1964).


\bibitem{spcond1}
Mathur, N.D., Grosche, F.M., Julian, S. R., Walker, I. R., Freye, D. M., Haselwimmer, R. K. W. \& Lonzarich, G. G. Magnetically mediated superconductivity in heavy fermion compounds, $Nature$ $(London)$ \textbf{394}, 39 (1998).
 
 
\bibitem{spcond2}
Yuan, H. Q., Grosche, F. M., Deppe, M., Geibel, C., Sparn, G. \& Steglich, F. Observation of two distinct superconducting phases in CeCu$_2$Si$_2$, $Science$  \textbf{302}, 2104 (2003). 

\bibitem{spcond3}
Mito, T., Kawasaki, S., Kawasaki, Y., Zheng, G.-q., Kitaoka, Y., Aoki, D., Haga, Y. \&  $\bar{\rm{O}}$nuki, Y. Coexistence of antiferromagnetism and superconductivity near the quantum criticality of the heavy-fermion compound CeRhIn$_5$, $Phys.$ $Rev.$ $Lett.$ \textbf{90}, 077004 (2003). 

\bibitem{spcond4}
Pham, L. D., Park, T., Maquilon, S., Thompson, J. D. \& Fisk, Z. Reversible tuning of the heavy-fermion ground state in CeCoIn$_5$. \textit{Phys.\ Rev.\ Lett.} \textbf{97}, 056404 (2006).


\bibitem{nonFermi}
Stewart, G. R. Non-Fermi-liquid behavior in $d$- and $f$-electron metals. \textit{Rev.\ Mod.\ Phys.} \textbf{73}, 797 (2001).


\bibitem{Doniach}
Doniach, S. The Kondo lattice and weak antiferromagnetism. \textit{Physica\ B} \textbf{91}, 231 (1977).


\bibitem{Doni1}
Tsunetsugu, H., Sigrist, M. \& Ueda, K. The ground-state phase diagram of the one-dimensional Kondo lattice model. \textit{Rev.\ Mod.\ Phys.} \textbf{69}, 809 (1997).


\bibitem{Doni2}
L$\ddot{\rm{o}}$hneysen, H., Rosch, A., Vojta, M. \& W$\ddot{\rm{o}}$lfle, P. Fermi-liquid instabilities at magnetic quantum phase transitions. \textit{Rev.\ Mod.\ Phys.} \textbf{79}, 1015 (2007).


\bibitem{Doni3}
Si, Q. \& Steglich, F. Heavy Fermions and quantum phase transitions. \textit{Science} \textbf{329}, 1161 (2010).



\bibitem{Kondo1}
Gegenwart, P., Si, Q. \& Steglich, F. Quantum criticality in heavy-fermion metals. \textit{Nat.\ Phys.} \textbf{4}, 186 (2008).


\bibitem{Kondo2}
Paschen, S. \& Si, Q. Quantum phases driven by strong correlations. \textit{Nat.\ Rev.\ Phys.} \textbf{3}, 9 (2021).



\bibitem{Kondo3}
Hafner, D. \textit{et al.} Possible quadrupole density wave in the superconducting Kondo lattice CeRh$_2$As. \textit{Phys.\ Rev.\ X} \textbf{12}, 011023 (2022).



\bibitem{1DKN1}
Jullien, R., Fields, J. N. \& Doniach, S. Zero-temperature real-space renormalization-group method for a Kondo-lattice model Hamiltonian. \textit{Phys.\ Rev.\ B} \textbf{16}, 4889 (1977).


\bibitem{1DKN2}
Moukouri, S., Caron, L. G., Bourbonnais, C. \& Hubert, L. Real-space density-matrix renormalization-group study of the Kondo necklace. \textit{Phys.\ Rev.\ B} \textbf{51}, 15920 (1995).


\bibitem{1DKN3}
Otsuka, H. \& Nishino, T. Gap-formation mechanism of the Kondo-necklace model. \textit{Phys.\ Rev.\ B} \textbf{52}, 15066 (1995).


\bibitem{1DKN4}
Zhang, G.-M., Gu, Q. \& Yu, L. Kondo spin liquid and magnetically long-range ordered states in the Kondo necklace model. \textit{Phys.\ Rev.\ B} \textbf{62}, 69 (2000).

\bibitem{1DKN5}
Essler, F. H. L., Kuzmenko, T. \& Zaliznyak, A. Luttinger liquid coupled to quantum spins: Flow equation approach to the Kondo necklace model. \textit{Phys.\ Rev.\ B} \textbf{76}, 115108 (2007).

\bibitem{1DKN6}
Mendoza-Arenas, J. J., Franco, R. \& Silva-Valencia, J. Gap formation and phase transition of the anisotropic Kondo necklace model: Density matrix renormalization group analysis. \textit{Phys.\ Rev.\ B} \textbf{81}, 035103 (2010).

\bibitem{myKondo}
Yamaguchi, H. \textit{et al.} Realization of a spin-1/2 Kondo necklace model with magnetic field-induced coupling switch. \textit{Phys.\ Rev.\ Res.} \textbf{7}, L012023 (2025).

\bibitem{SPT}
Hidaka, Y., Furuya, S. C., Ueda, A. \& Tada, Y. Gapless symmetry-protected topological phase of quantum antiferromagnets on anisotropic triangular strip. \textit{Phys.\ Rev.\ B} \textbf{106}, 144436 (2022).

\bibitem{TominagaCo}
Yamaguchi, H. \textit{et al.} A ladder-based 2D spin model in a radical-Co complex. \textit{Phys.\ Rev.\ B} \textbf{107}, 174422 (2023).

\bibitem{TominagaNi}
Tominaga, Y. \textit{et al.} Mixed-spin two-dimensional lattice composed of spins 1/2 and 1 in a radical-Ni complex. \textit{Phys.\ Rev.\ B} \textbf{108}, 024424 (2023).








\bibitem{neel}
Chen, W., Hida, K. \& Sanctuary, B. C. Ground-state phase diagram of $S$ = 1 XXZ chains with uniaxial single-ion-type anisotropy. \textit{Phys.\ Rev.\ B} \textbf{67}, 104401 (2003).



\bibitem{peak}
Johnston, D. C. \textit{et al.} Thermodynamic properties of low-dimensional quantum spin systems with energy gaps. \textit{Phys.\ Rev.\ B} \textbf{61}, 9558 (2000).



\bibitem{morotaMn}
Yamaguchi, H. \textit{et al.} Observation of thermodynamics originating from a mixed-spin ferromagnetic chain. \textit{Phys.\ Rev.\ B} \textbf{106}, L100404 (2022).


\bibitem{MOcal} 
Shoji, M. \textit{et al.} A general algorithm for calculation of Heisenberg exchange integrals $J$ in multispin systems. \textit{Chem.\ Phys.\ Lett.} \textbf{432}, 343--347 (2006).


\bibitem{QMC2} 
Sandvik, A. W. Stochastic series expansion method with operator-loop update. \textit{Phys.\ Rev.\ B} \textbf{59}, R14157--R14160 (1999).


\bibitem{ALPS} 
Albuquerque, A. F. \textit{et al.} The ALPS project release 1.3: Open-source software for strongly correlated systems. \textit{J.\ Magn.\ Magn.\ Mater.} \textbf{310}, 1187--1193 (2007).


\bibitem{ALPS3} 
Bauer, B. \textit{et al.} The ALPS project release 2.0: open source software for strongly correlated systems. \textit{J.\ Stat.\ Mech.} \textbf{2011}, P05001 (2011).


\end{thebibliography}
\end{document}